\documentclass[
aps,
pra,
superscriptaddress,
nofootinbib,
twocolumn
]{revtex4-2}

\usepackage{amsmath}
\allowdisplaybreaks[4]
\usepackage{amssymb}
\usepackage{amsfonts}
\usepackage[mathscr]{euscript}

\usepackage{graphicx}
\usepackage{xcolor}
\usepackage{qcircuit}

\usepackage{bm}
\usepackage{dcolumn}
\usepackage[normalem]{ulem}
\usepackage{rotating}
\usepackage{tikz}

\usepackage[colorlinks=true,linkcolor=magenta,citecolor=blue]{hyperref}

\newcommand{\Tr}{\mathrm{Tr}}

\newcommand{\abs}[1]{\vert #1 \vert}
\newcommand{\absLR}[1]{\left\vert #1 \right\vert}

\newcommand{\ket}[1]{\vert #1 \rangle}
\newcommand{\bra}[1]{\langle #1 \vert}
\newcommand{\ketbra}[2]{\vert #1 \rangle \langle #2 \vert}

\newcommand{\mean}[1]{\langle #1 \rangle}

\begin{document}

\title{Quantum algorithms for Schrieffer–Wolff transformation}

\author{Zongkang Zhang}
\affiliation{Graduate School of China Academy of Engineering Physics, Beijing 100193, China}

\author{Yongdan Yang}
\affiliation{Graduate School of China Academy of Engineering Physics, Beijing 100193, China}

\author{Xiaosi Xu}
\email{xsxu@gscaep.ac.cn}
\affiliation{Graduate School of China Academy of Engineering Physics, Beijing 100193, China}

\author{Ying Li}
\email{yli@gscaep.ac.cn}
\affiliation{Graduate School of China Academy of Engineering Physics, Beijing 100193, China}
\date{\today}

\begin{abstract}
The Schrieffer–Wolff transformation aims to solve degenerate perturbation problems and give an effective Hamiltonian that describes the low-energy dynamics of the exact Hamiltonian in the low-energy subspace of unperturbed Hamiltonian. This unitary transformation decoupling the low-energy and high-energy subspaces for the transformed Hamiltonian can be realized by quantum circuits. We give a fully quantum algorithm for realizing the SW transformation. We also propose a hybrid quantum-classical algorithm for finding the effective Hamiltonian on NISQ hardware, where a general cost function is used to indicate the decoupling degree. Numerical simulations without or with noise and experiments on quantum computer {\it ibmq\_manila} are implemented for a Heisenberg chain model. Our results verify the algorithm and show its validity on near-term quantum computers. 
\end{abstract}

\maketitle

\section{Introduction}

Understanding and controlling quantum many-body systems is of crucial importance in modern physics \cite{vidal2004efficient,islam2015measuring,carleo2017solving}. 
With the exponential growth of Hilbert space, it is difficult to perform accurate analytical calculations as the system size increases. 
In most applications, a description about the low-energy properties is sufficient, thus the study of the low-energy effective Hamiltonian ($H_{\rm eff}$) plays an important role in many-body physics \cite{soliverez1981general}. 
The Schrieffer–Wolff transformation (SWT) was originally proposed in Ref.~\cite{schrieffer1966relation}, where the Kondo model is obtained from the Anderson impurity model in the strong coupling regime with a unitary transformation. 
SWT extracts $H_{\rm eff}$ from the exact Hamiltonian by decoupling the low-energy and high-energy subspaces. It is regarded as the operator version of the degenerate perturbation theory \cite{bravyi2011schrieffer}. For instance, with SWT one can infer that the Fermi-Hubbard model is equivalent to Heisenberg model in the strong coupling limit \cite{macdonald1988t,fazekas1999lecture}, where the perturbation theory approach becomes impractical \cite{cleveland1976obtaining}. Being widely applied and developed in many kinds of quantum problems \cite{paaske2005vibrational,issler2010nuclear,hohenester2010cavity,erlingsson2010energy,uchoa2011kondo,kessler2012generalized,heikkila2014enhancing,bukov2015universal,bukov2016schrieffer,zhang2017rescuing,matlack2018designing,yan2018tunable,wurtz2020variational,garbe2020critical,murakami2021high}, SWT also has different names across various fields: Foldy–Wouthuysen transformation in relativistic quantum mechanics \cite{foldy1950dirac}, Fr{\"o}hlich transformation in electron-phonon interaction \cite{frohlich1952interaction} and $k\cdot p$ method in semiconductor physics \cite{winkler2003spin}.  

SWT provides a systematic perturbative method for computing $H_{\rm eff}$ at any order \cite{bravyi2011schrieffer}. The transformation is preformed by a unitary operator defined as $U=e^S$, where $S$ is an anti-Hermitian operator and is called the generator. The coefficients in the Taylor series of $S$ can be derived order by order through an inductive formula, which is hard for high orders. What's more, the calculation is carried out in the space of the whole system, which scales up exponentially and thus makes such procedure impractical on classical computers for large systems. In most applications, only the second-order $H_{\rm eff}$ \cite{wagner1986unitary} is considered. 

Quantum computation has attracted much attention in the recent years. It has been proved that quantum computers are capable of handling problems which are intractable for classical computers \cite{shor1994algorithms,arute2019quantum}. 
While the fault-tolerant universal quantum computer \cite{shor1996fault,gottesman1998theory} may not be feasible in near term, noisy intermediate-scale quantum (NISQ) computers are possible candidates for applications in various fields including many-body quantum physics \cite{preskill2018quantum}.
Starting with the Variational Quantum Eigensolver (VQE)~\cite{peruzzo2014variational}, which is a Variational Quantum Algorithm (VQA)~\cite{cerezo2021variational} to estimate the ground state of a given Hamiltonian using shallow circuits, many useful quantum algorithms for NISQ computers have been proposed \cite{kandala2017hardware,li2017efficient,farhi2018classification,xu2021variational}.

In this paper, we for the first time propose two quantum algorithms to realize SWT and obtaining the effective Hamiltonian. The first algorithm constructs the unitary $U$ directly using a quantum circuit under the observation that $U$ is composed of two reflection operators acting on the unperturbed and perturbed Hamiltonians. The reflection operators effectively apply a phase flip to the high-energy eigenstates and $U$ additionally applies a conditional phase change to the state. This method takes use of the quantum phase estimation algorithm to distinguish the states corresponding to low- and high-energy eigenspaces, and thus may require circuit depths exceeding the limits of hardware available in the NISQ era. Therefore, we propose the second algorithm for NISQ devices. It is a hybrid quantum-classical algorithm based on the variational algorithm. The cost function is designed such that it can reach the minimum when the evolution of a state within the low-energy subspace of the unperturbed Hamiltonian is independent of time. The parameterized cost function can be measured using a quantum computer, and the parameters are to be optimized with a classical algorithm. To demonstrate the effectiveness of our second algorithm, we performed numerical simulations using QuESTlink \cite{jones2020questlink} and {\it qasm\_simulator} as well as experiments on the IBMQ quantum computer {\it ibmq\_manila}, with the example of a four-qubit Heisenberg chain with long-distance entanglement between the two ends.

This paper is organized as follows. In section~\ref{SWT}, we give a brief introduction on SWT. Next, in section~\ref{first_algorithm} and section~\ref{Methods} we describe the two quantum algorithms to construct SWT respectively. The proof-of-principle numerical simulations and experiments are presented in section~\ref{Results}. This paper ends with a conclusion in section~\ref{Conclusion}.

\section{\label{SWT}Schrieffer–Wolff transformation}
In degenerate perturbation theory, we consider a Hamiltonian $H=H_0+\epsilon V$, where $H_0$ is the unperturbed Hamiltonian with well-separated low- and high-energy levels, and $\epsilon V$ is the perturbation which brings to split of the spectra. The Schrieffer–Wolff transformation (SWT) is a unitary transformation that maps the total Hamiltonian $H$ to the low-energy effective Hamiltonian $H_{\rm eff}$, which acts on the low-energy subspace of $H_0$ and reproduces the low-energy spectrum of $H$.

The unperturbed Hamiltonian $H_0$ can be written into low energy and high energy levels according to spectral decomposition
\begin{eqnarray}
H_0 = \sum_{i=1}^M E_i^{(0)} \ket{\phi_i^{(0)}}\bra{\phi_i^{(0)}} + \sum_{i=M+1}^N E_i^{(0)} \ket{\phi_i^{(0)}}\bra{\phi_i^{(0)}},
\end{eqnarray}
with eigenvalues $E_1^{(0)} \leq \cdots \leq E_M^{(0)} < E_{M+1}^{(0)} \leq \cdots \leq E_N^{(0)}$. 
Here $N$ is the dimension of system's Hilbert space and $M$ is the dimension of the low-energy eigenspace of $H_0$, denoted as $\mathscr{P}_0$.
Define $P_0$ as the projector on $\mathscr{P}_0$ 
\begin{eqnarray}
P_0 = \sum_{i=1}^M \ket{\phi_i^{(0)}}\bra{\phi_i^{(0)}},
\end{eqnarray}
and $Q_0 = I - P_0$ is its complement. 

The low-energy and high-energy spectra are separated by the gap
\begin{eqnarray}
\Delta = E_{M+1}^{(0)} - E_M^{(0)}.
\end{eqnarray}
It is assumed that $\epsilon V$ is small and satisfies 
\begin{eqnarray}
\Vert \epsilon V \Vert < \frac{\Delta}{2},
\label{eq:shift}
\end{eqnarray}
where $\Vert \cdot \Vert$ is the operator norm.
Since the perturbation shifts the eigenvalues of $H_0$ by at most $\Vert \epsilon V \Vert $, there will still have a positive gap between the low-energy and the high-energy spectra.

So the total Hamiltonian can be written as 
\begin{eqnarray}
H = \sum_{i=1}^M E_i \ket{\phi_i}\bra{\phi_i} + \sum_{i=M+1}^N E_i \ket{\phi_i}\bra{\phi_i},
\end{eqnarray}
with its eigenvalues $E_1 \leq \cdots \leq E_M < E_{M+1} \leq \cdots \leq E_N$. 
Denote $\mathscr{P}$ as the M-dimension low-energy eigenspace of $H$ and define $P$ as the projector on $\mathscr{P}$, then $Q = I - P$ is its complement. Thus the total Hamiltonian can be expressed as a block-diagonal form
\begin{eqnarray}
H = P H P + Q H Q.
\label{eq:block}
\end{eqnarray}
$P$ and $P_0$ can be connected with a unitary $U$
\begin{eqnarray}
U P U^\dag=P_0,
\label{eq:mapP}
\end{eqnarray}
which is exactly the Schrieffer–Wolff transformation. 
Likewise, we can find
\begin{eqnarray}
U Q U^\dag=Q_0.
\label{eq:mapQ}
\end{eqnarray}

It has been proved in~\cite{bravyi2011schrieffer} that $U$ can be constructed by reflection operators $R_{\mathscr{P}_0}$ and $R_{\mathscr{P}}$
\begin{eqnarray}
U=\sqrt{R_{\mathscr{P}_0}R_{\mathscr{P}}},
\label{eq:def}
\end{eqnarray}
where $R_{\mathscr{P}_0} = 2P_0 - I$ and $R_{\mathscr{P}} = 2P - I$. Note that in some places $U$ is defined as $U=e^S$, where $S$ is called the generator of the transformation~\cite{haq2020systematic, kessler2012generalized}.

Using Eq.(\ref{eq:block}), Eq.(\ref{eq:mapP}) and Eq.(\ref{eq:mapQ}), the transformed Hamiltonian $H^\prime$ can be reexpressed as \cite{consani2020effective}
\begin{align}
H^\prime &=U H U^\dag\\
&= U P H P U^\dag + U Q H Q U^\dag\\
&=P_0 U H U^\dag P_0 + Q_0 U H U^\dag Q_0,
\end{align}
which indicates that $H^\prime$ is block-diagonal with respect to $P_0$ and $Q_0$. Project $H^\prime$ onto $\mathscr{P}_0$ yields the low-energy effective Hamiltonian
\begin{eqnarray}
H_{\rm eff}=P_0UHU^\dag P_0,
\label{eq:Heff}
\end{eqnarray}
whose $M$ eigenvalues are the same as the $M$ lowest eigenvalues of $H$. 

To obtain $H_{\rm eff}$, one needs to find $U$. In the following, we propose two quantum algorithms to find $H_{\rm eff}$. The first algorithm is for universal fault-tolerant quantum computers and the second one is a hybrid approach based on the variational approach and thus is suitable for near-term quantum devices.

\section{\label{first_algorithm}The quantum algorithm for SWT}

In many models, the Hamiltonian $H$ can be decomposed into Pauli terms
\begin{eqnarray}
H=\sum_{i=1}^{N_h} h_i \sigma^i,
\end{eqnarray}
and the term number increases polynomially with the system size, e.g.~the Heisenberg model and Fermi-Hubbard model~\cite{arute2020observation}.
For each $\sigma_i$, we can construct a density matrix $\rho_i=(\sigma_i+\openone)/2^n$, thus $\sigma_i = 2^n\rho_i-\openone$. Then, the effective Hamiltonian in Eq.(\ref{eq:Heff}) becomes
\begin{eqnarray}
H_{\rm eff}=\sum_{i=1}^{N_h} h_i \left(2^nP_0U \rho_i U^\dag P_0 - P_0\right).
\end{eqnarray}
Notice that $\rho_i$ is a quantum state which can be prepared with a quantum computer, the idea is to apply $U$ to the state, perform projective measurement onto $P_0$ and then obtain $P_0 U \rho_i U^\dag P_0$ (by post-selection according to the measurement result). The full tomography of each $P_0 U \rho_i U^\dag P_0$ is unrealistic when the system size is large, therefore, we need an ansatz of the effective Hamiltonian $H_{\rm eff} = \sum_{\tau\in A} g_\tau \tau$. Here $A$ is a subset of Pauli operators, and we can generate the subset from Pauli terms in $H$. Then,
\begin{eqnarray}
g_\tau &=& \sum_{i=1}^{N_h} h_i \left[\Tr(\tau P_0U \rho_i U^\dag P_0) - 2^{-n}\Tr(\tau P_0)\right],
\end{eqnarray}
where $n$ is the qubit number. In general, we can realise the projective measurement onto $P_0$ using QPE: We use QPE to measure the eigenenergy of the Hamiltonian, but the final measurement is adapted for distinguishing two subspaces $P_0$ and $Q_0$ instead of specific eigenenergies. The total number expected values evaluated for constructing $H_{\rm eff}$ is $N_h\times\abs{A}$. Usually $N_h$ increases polynomially with the system size. Therefore, by taking an ansatz in which $\abs{A}$ is ploynomial, the overall cost for reconstructing $H_{\rm eff}$ is polynomial. Next, we show that $U$ is equivalent to an oracle which does conditional phase rotations to some states. 

Given an arbitrary state, it can be written in the basis of the eigenspace of $H$
\begin{eqnarray}
\ket{\Psi}=\sum_{i=1}^{N}\alpha_i \ket{\phi_i},
\end{eqnarray}
thus $e^{iH}\ket{\Psi}=\sum_{i=1}^{N}\alpha_i e^{iE_i}\ket{\phi_i}$.
As $R_{\mathscr{P}}=2P-I$ and $P$ is the projector on the low-energy eigenspace, $R_{\mathscr{P}}$ is effectively a reflection operator that does a phase flip to the eigenstates corresponding to the high-energy space, namely
\begin{eqnarray}
R_{\mathscr{P}}\ket{\Psi} = \sum_{i=1}^{M}\alpha_i \ket{\phi_i}
-\sum_{i=M+1}^{N}\alpha_i \ket{\phi_i}.
\label{eq:rp}
\end{eqnarray}
The same applies for $R_{\mathscr{P}_0}$.

On a quantum computer, we could use QPE to realize the unitary operator $R_{\mathscr{P}}$,
\begin{eqnarray}
&&\ket{+}^{\otimes l}\otimes\ket{\Psi}=\frac{1}{\sqrt{2^l}}\sum_{x=0}^{2^l-1}\ket{x}\otimes\sum_{i=1}^{N}\alpha_i \ket{\phi_i}\notag\\
&\stackrel{U_{QPE}}{\longrightarrow}& \sum_{i=1}^N\alpha_i\sum_{k=0}^{2^l-1}f_{i}(k)\ket{k}\otimes\ket{\phi_i},
\label{eq:CirRp}
\end{eqnarray}
where $l$ is the number of ancilla qubits, $U_{QPE} = \sum_{x=0}^{2^l-1} \ketbra{x}{x}\otimes e^{-iHtx}$, $t$ is chosen such that eigenvalues of $Ht$ are within the interval $[0,2\pi)$, and $f_{i}(k) = \frac{1}{2^l}\sum_{x=0}^{2^l-1}e^{i(\frac{2\pi k}{2^l}-E_it)x}$. 
As the amplitude $f_{i}(k)$ is concentrated at $k\simeq \frac{2^l E_it}{2\pi}$, we could apply a phase flip to the states with $E_i> E_{M+1}^{(0)} - \Delta/2$, by apply the unitary gate $\sum_{k=0}^{k_{th}-1}\ketbra{k}{k} - \sum_{k=k_{th}}^{2^l-1}\ketbra{k}{k}$ on ancilla qubits, where $k_{th} = \left\lceil \frac{2^l (E_{M+1}^{(0)} - \Delta/2)t}{2\pi} \right\rceil$. Then we apply $U_{QPE}^{-1}$ to get
\begin{eqnarray}
\frac{1}{\sqrt{2^l}}\sum_{x=0}^{2^l-1}\ket{x}\otimes \left(\sum_{i=1}^{M}\alpha_i \ket{\phi_i}-\sum_{i=M+1}^{N}\alpha_i \ket{\phi_i}\right),
\end{eqnarray}
up to a small error due to the finite resolution of QPE (i.e.~$2^l$ is finite). 
Remove the ancilla qubits, we get Eq.(\ref{eq:rp}). We then apply $R_{\mathscr{P}_0}$ to $R_{\mathscr{P}}\ket{\Psi}$ to get $R_{\mathscr{P}_0}R_{\mathscr{P}}\ket{\Psi}$.

As $U=\sqrt{R_{\mathscr{P}_0}R_{\mathscr{P}}}$, if eigenvalues and eigenvectors of $R_{\mathscr{P}_0}R_{\mathscr{P}}$ are $e^{i\theta_j}$ and $\ket{\psi_j}$, i.e.
\begin{eqnarray}
R_{\mathscr{P}_0}R_{\mathscr{P}}\ket{\Psi}=\sum_{j}\beta_j e^{i\theta_j}\ket{\psi_j},
\end{eqnarray}
we have
\begin{eqnarray}
U\ket{\Psi}=\sum_{j}\beta_j e^{i\frac{\theta_j}{2}}\ket{\psi_j}.
\end{eqnarray}
Therefore, we could use QPE to realise the phase $e^{i\frac{\theta_j}{2}}$:
\begin{eqnarray}
&&\ket{+}^{\otimes l}\otimes\ket{\Psi}=\frac{1}{\sqrt{2^l}}\sum_{x=0}^{2^l-1}\ket{x}\otimes\sum_{j}\beta_j\ket{\psi_j}\notag\\
&\stackrel{U_{QPE}^{\prime}}{\longrightarrow}& \sum_{j}\beta_j\sum_{k=0}^{2^l-1}g_{j}(k)\ket{k}\otimes\ket{\psi_j},
\label{eq:CirU}
\end{eqnarray}
where $U_{QPE}^{\prime} = \sum_{x=0}^{2^l-1} \ketbra{x}{x}\otimes (R_{\mathscr{P}_0}R_{\mathscr{P}})^x$ and $g_{j}(k) = \frac{1}{2^l}\sum_{x=0}^{2^l-1}e^{i(\frac{2\pi k}{2^l}+\theta_j)x}$. 
Because $g_{j}(k)$ is concentrated at $k\simeq -\frac{2^l \theta_j}{2\pi}$, the phase $\theta_i$ is (approximately) stored on the ancilla qubits. We then apply the phase gate $\sum_{k=0}^{2^l-1}e^{-i\frac{\pi k}{2^l}}\ketbra{k}{k}$, before applying $U_{QPE}^{\prime -1}$. Therefore, the overall effect of all these steps applies a phase of $e^{i\frac{\theta_i}{2}}$ to the state $\ket{\psi_i}$ and thus effectively realizes $U\ket{\Psi}$.
All the details are given in Appendix~\ref{Appendix details}.

As this algorithm needs to apply a concatenated QPE, which uses deep circuits and a reasonable number of ancilla qubits for high resolutions of energy and phase, it is not feasible until large-scale error-protected quantum computers come to exist. In the next section, we introduce a hybrid algorithm which is suitable for near-term quantum devices. 

\section{\label{Methods}The hybrid quantum-classical algorithm for SWT}

If $UHU^\dagger$ is block-diagonal with respect to $P_0$ and $Q_0$, we can infer that $H$ is block-diagonal with respect to $P$ and $Q$, where $P = U^\dag P_0 U$ and $Q = U^\dag Q_0 U$ are projectors transformed by inverse SWT. Under this condition, a state within the subspace of $\mathscr{P}$ should remain in the subspace as it evolves. Therefore, we can design a cost function as the following
\begin{eqnarray} 
\mathcal{L}_t(\vec{\theta})=-\frac{1}{M}\sum_{i,j}^M\left|\bra{\phi_i^{(0)}}U(\vec{\theta}) e^{-iHt} U^\dag(\vec{\theta})\ket{\phi_j^{(0)}}\right|^2,
\label{eq:evolution}
\end{eqnarray}
where $\ket{\phi_i^{(0)}}$ and $\ket{\phi_j^{(0)}}$ are the basis states in $\mathscr{P}_0$. We remark that we can use the Monte Carlo method to evaluate the cost function rather than measure each term in it. We find the global minimum of the cost function is $-1$, obtained when $U(\vec{\theta})HU^\dagger(\vec{\theta})$ is block-diagonal with respect to $P_0$ and $Q_0$. However, to minimize this cost function using a quantum computer, one needs to implement $e^{-iHt}$, which usually requires deep circuits to achieve high accuracy. 

Now, we introduce an alternative cost function. When $U(\vec{\theta})HU^\dagger(\vec{\theta})$ is block-diagonal with respect to $P_0$ and $Q_0$, the minimum cost $\mathcal{L}_{t,min}$ should be invariant when the time $t$ changes. Therefore, we rewrite $\mathcal{L}_t$ in the form of Taylor series:
\begin{eqnarray}
\mathcal{L}_t=A+Bt+Ct^2+Dt^3+\cdots.
\end{eqnarray}
Extract the coefficients $A$, $B$ and $C$, we find
\begin{eqnarray}
A&=&-1\\
B&=&0\\
C&=&\frac{1}{M}\left(\sum_{i=1}^M\bra{\phi_i^{(0)}}U(\vec{\theta})H^2U^\dag(\vec{\theta})\ket{\phi_i^{(0)}}\right. \notag\\
&-&\left.\sum_{i, j}^M\absLR{ \bra{\phi_i^{(0)}}U(\vec{\theta})HU^\dag(\vec{\theta})\ket{\phi_j^{(0)}} }^2\right).
\end{eqnarray}
The derivation is given in Appendix~\ref{Appendix derivation}. Because $\mathcal{L}_{t,min}$ is time-independent, $C$ should be $0$ when $\mathcal{L}_t$ reaches minimum.

In fact, we find that $C=0$ is equivalent to block diagonalization of $H$ with respect to $P$ and $Q$. See Appendix~\ref{Appendix equivalence} for details. With this we can reconstruct the cost function as the absolute value of $C$
\begin{eqnarray}
\mathcal{L}(\vec{\theta})&=&\abs{C} \notag\\
&=&\frac{1}{M}\left\vert \sum_{i=1}^M\bra{\phi_i^{(0)}}U(\vec{\theta})H^2U^\dag(\vec{\theta})\ket{\phi_i^{(0)}}\right. \notag\\
&-&\left.\sum_{i, j}^M\absLR{ \bra{\phi_i^{(0)}}U(\vec{\theta})HU^\dag(\vec{\theta})\ket{\phi_j^{(0)}} }^2 \right\vert,
\label{eq:loss}
\end{eqnarray}
which has the global minimum $0$. Note that we make $\mathcal{L}(\vec{\theta})$ the absolute value of $C$ to avoid the case where it becomes negative with experimental noise.

Now we have the final version of the cost function. Starting from an initial parameter set $\vec{\theta_0}$, our hybrid algorithm optimizes the set $\vec{\theta}$ such as $\mathcal{L}(\vec{\theta})$ reaches minimum after several iteration cycles. In each cycle, $\mathcal{L}(\vec{\theta})$ is measured using a quantum computer, and a classical algorithm is used to optimize $\vec{\theta}$ based on $\mathcal{L}(\vec{\theta})$. The iteration continues until $\mathcal{L}(\vec{\theta})$ reaches its minimum. Then we obtain the corresponding $U(\vec{\theta})$ as a good approximation of $U$. 

To implement the hybrid algorithm, the basis set \{$\ket{\phi_i^{(0)}}$\} of the unperturbed Hamiltonian $H_0$ must be known. $U(\vec{\theta})$ is constructed by a parameterized quantum circuit. The Hamiltonian $H$ is decomposed into Pauli terms $H=\sum_l n_l\sigma^l$, thus each term in $\mathcal{L}(\vec{\theta})$ has the form $\bra{\phi_i^{(0)}}U(\vec{\theta})\sigma^lU^\dag(\vec{\theta})\ket{\phi_j^{(0)}}$, which can be measured using a quantum computer \cite{mitarai2019methodology}. In the following section, we demonstrate the hybrid algorithm using an example of a one-dimensional Heisenberg model. 

\section{\label{Results}Simulation and Experiment}
We demonstrate the effectiveness of the hybrid algorithm with experiments on an IBMQ quantum device. We choose a spin model to verify our methods in Section~\ref{Methods}. The system is an antiferromagnetic Heisenberg chain with modulated interaction strengths~\cite{li2005quantum}. Two spins at the ends of the chain are weakly coupled to other spins on the chain. If the chain with two ends removed has a non-degenerate ground state and an energy gap above the ground state, two end spins are effectively coupled through the chain: according to the perturbation theory, two end spins are directly coupled in the effective model. We will reconstruct the effective model of the Heisenberg chain with experiments on IBMQ devices.

\subsection{Model}
The unperturbed Hamiltonian is the chain with two end spins decoupled, i.e. 
\begin{eqnarray}
H_0&=&2\sum_{i=2}^{N-2}(\sigma^x_i\sigma^x_{i+1}+\sigma^y_i\sigma^y_{i+1}+\sigma^z_i\sigma^z_{i+1}),
\end{eqnarray}
where $N$ is the total number of spins. The perturbation is the interaction between end spins and the rest of the chain, i.e. 
\begin{eqnarray}
V&=&\sigma^x_1\sigma^x_2+\sigma^y_1\sigma^y_2+\sigma^z_1\sigma^z_2 \notag\\
&+&\sigma^x_{N-1}\sigma^x_N+\sigma^y_{N-1}\sigma^y_N+\sigma^z_{N-1}\sigma^z_N.
\end{eqnarray}

Thinking of that end spins $1$ and $N$ are removed from the system, the Hamiltonian of spins $2$ to $N-1$ is $H_0$. Then, if the ground state of $H_0$ (without considering spins $1$ and $N$) is non-degenerate, the unperturbed ground-state subspace of all spins is four-fold degenerate. Let $\ket{GS}$ be a state of spins $2$ to $N-1$ and the non-degenerate ground state of $H_0$, the ground-state subspace of all spins has the basis $\left\{ \ket{\mu}\otimes\ket{GS}\otimes\ket{\nu}~\vert~ \mu,\nu=0,1\right\}$. If $N=4$, the subsystem ground state is the singlet state $\ket{GS} = \frac{1}{\sqrt{2}}(\ket{0}\otimes\ket{1}-\ket{1}\otimes\ket{0})$; in general, we can obtain the ground state $\ket{GS}$ via VQE. 

With perturbation, usually the ground state splits into singlet ground state and triplet excited states because of the symmetry of the Heisenberg interaction. So the effective Hamiltonian acting in this low-energy subspace is equivalent to Heisenberg interaction between spins $1$ and $N$. The effective ground state of the subspace of spins $1$ and $N$ is then $\frac{1}{\sqrt{2}}(\ket{01}_{1,N}-\ket{10}_{1,N})$. The entanglement between two separated spins comes from repeated nearest-neighbor interactions in the Heisenberg chain.

\subsection{Ansatz}

Here we adopt an empirical ansatz to construct the transformation $U$. If $H_0$ and $V$ commute, they are simultaneously (block-) diagonalizable by a unitary transformation, and $U$ is identity. So this commutative situation is trivial. If $H_0$ and $V$ do not commute, we can express the commutator as a linear combination of Pauli operators: 
\begin{eqnarray}
[H_0,V]=i\sum_{j=1}^M \alpha_j \sigma^{(j)},
\label{eq:hvcommute}
\end{eqnarray}
where $\alpha_i$ is a real scalar coefficient. Note that usually the number of Pauli operators $M$ in the linear combination increases polynomially with the system size. Given the commutator, we approximate the generator $S$ with an operator in the form~\cite{haq2020systematic}  
\begin{eqnarray}
\eta = \sum_{j=1}^M \eta_j \sigma^{(j)}.
\end{eqnarray}
Suppose coefficients $\eta_j$ are small, we can approximate $U = e^S$ with the ansatz transformation
\begin{eqnarray}
U(\vec{\theta})=\prod_{j=1}e^{i \sigma^{(j)} \theta_j},
\end{eqnarray}
where parameters $\theta_j$ are optimized according to our algorithm in Section~\ref{Methods}. 

Taking $N = 4$ in the model, we find that $U$ has 12 terms. By removing parameters that have little impact on the cost function, $U$ is simplified to 6 terms to reduce circuit depth. Considering symmetry of the system, we keep three parameters. Finally, the ansatz is
\begin{eqnarray}
U(\vec{\theta})&=&e^{i \sigma^z_2\sigma^y_3\sigma^x_4\theta_1/2}e^{i \sigma^z_2\sigma^x_3\sigma^y_4\theta_2/2}e^{i \sigma^y_2\sigma^x_3\sigma^z_4\theta_3/2} \notag\\
&\,&e^{i \sigma^z_1\sigma^x_2\sigma^y_3\theta_3/2}e^{i \sigma^y_1\sigma^x_2\sigma^z_3\theta_2/2}e^{i \sigma^x_1\sigma^y_2\sigma^z_3\theta_1/2},
\end{eqnarray}
and the corresponding circuit is drawn in Appendix~\ref{Appendix details}.

\subsection{Measurement circuits}

As mentioned before, one needs to measure the transition amplitude $\bra{\phi_i^{(0)}}U(\vec{\theta})\sigma^lU^\dag(\vec{\theta})\ket{\phi_j^{(0)}}$, where $\ket{\phi_j^{(0)}}$ is one of the four basis states of the ground-state subspace. We can measure the transition amplitude using a Hadamard test circuit~\cite{kitaev1995quantum}. To minimise the gate number, which is important on NISQ devices, we measure the transition amplitude in the following way. We consider two cases. In the first case, when $i=j$, the transition amplitude is the expected value of $\sigma^l$ in the state $U^\dag(\vec{\theta})\ket{\phi_j^{(0)}}$, which can be directly measured: We prepare the state $\ket{\phi_j^{(0)}}$, then apply the transformation $U^\dag(\vec{\theta})$ and finally measure $\sigma^l$. In the second case, when $i\neq j$, we always have $\ket{\phi_j^{(0)}} = G\ket{\phi_i^{(0)}}$, where $G = \openone,\sigma_1^x,\sigma_N^x,\sigma_1^x\sigma_N^x$. Then the transition amplitude becomes 
\begin{eqnarray}
&& \bra{\phi_i^{(0)}}U(\vec{\theta})\sigma^lU^\dag(\vec{\theta})\ket{\phi_j^{(0)}} \notag \\
&=& \bra{\phi_i^{(0)}}U(\vec{\theta})\sigma^lU^\dag(\vec{\theta})G\ket{\phi_i^{(0)}} \notag \\
&=&\bra{\phi_i^{(0)}}\frac{I+G}{2}U(\vec{\theta})\sigma^lU^\dag(\vec{\theta}) \frac{I+G}{2}\ket{\phi_i^{(0)}}\notag\\
&&-\bra{\phi_i^{(0)}}\frac{I-G}{2}U(\vec{\theta})\sigma^lU^\dag(\vec{\theta}) \frac{I-G}{2}\ket{\phi_i^{(0)}} \notag\\
&&-i\bra{\phi_i^{(0)}}\frac{I-iG}{2}U(\vec{\theta})\sigma^lU^\dag(\vec{\theta}) \frac{I+iG}{2}\ket{\phi_i^{(0)}}\notag\\
&&+i\bra{\phi_i^{(0)}}\frac{I+iG}{2}U(\vec{\theta})\sigma^lU^\dag(\vec{\theta}) \frac{I-iG}{2}\ket{\phi_i^{(0)}}.
\end{eqnarray}
According to the above equation, the transition amplitude becomes a linear combination of expected values of $\sigma^l$ in states $\frac{I\pm G}{2}\ket{\phi_i^{(0)}}$ and $\frac{I\pm iG}{2}\ket{\phi_i^{(0)}}$ \cite{mitarai2019methodology}. We can measure these expected values by preparing these four states (up to the normalization factor). In our case, because the transition amplitude is always real, we only need to prepare states $\frac{I\pm G}{2}\ket{\phi_i^{(0)}}$. See Table~\ref{tab:table1} for combinations of $\ket{\phi_i^{(0)}}$ and $G$. 

\begin{table}[h]
\begin{tabular}{|c|c|c|c|}
\hline
$\bra{\phi_i^{(0)}}$ & $\ket{\phi_j^{(0)}}$ & $G$ \\
\hline
$\bra{GS}$ & $\sigma^x_1\ket{GS}$ & $\sigma^x_1$  \\
\hline
$\bra{GS}$ & $\sigma^x_N\ket{GS}$ & $\sigma^x_N$ \\
\hline
$\bra{GS}$ & $\sigma^x_1\sigma^x_N\ket{GS}$ & $\sigma^x_1\sigma^x_N$ \\
\hline
$\bra{GS}\sigma^x_1$ & $\sigma^x_N\ket{GS}$ & $\sigma^x_1\sigma^x_N$ \\
\hline
$\bra{GS}\sigma^x_1$ & $\sigma^x_1\sigma^x_N\ket{GS}$ & $\sigma^x_N$ \\
\hline
$\bra{GS}\sigma^x_N$ & $\sigma^x_1\sigma^x_N\ket{GS}$ & $\sigma^x_1$ \\
\hline
\end{tabular}
\caption{\label{tab:table1} All possible combinations of $\ket{\phi_i^{(0)}}$ and $G$.}
\end{table}

\subsection{Implementation and Results}

Now we demonstrate our algorithm on an IBM quantum device ${\it ibmq\_manila}$, a five-qubit superconducting quantum computer with readout error rates $2.02\% \sim 3.53\%$, single-qubit gate error rates $0.02\% \sim 0.03\%$ and CNOT gate error rates $0.55\% \sim 1.23\%$ during our experiments. 

In each experiment, the initial parameters $\theta_i$ were set to zero, and the SPSA \cite{spall1998implementation} algorithm was used to optimize parameters in each cycle of the iteration. The optimization process continued until the cost function stops to decline in several successive iterations. During the experiment, the Clifford data regression (CDR) error mitigation technique~\cite{czarnik2021error}, which is a simplified version of Clifford data learning\cite{strikis2021learning}, was used to improve the result. Each circuit was ran for $10^{4}$ shots.

The amplitudes obtained in quantum computer were used to construct $H_{\rm eff}$. This is equivalent to the full tomography of each $U\sigma^i U^\dag$ in the subspace $P_0$. In this experiment, $H_{\rm eff}$ is a $4\times4$ matrix with the element in row $i$ and column $j$ given by $\bra{\phi_i^{(0)}}U(\vec{\theta})HU^\dag(\vec{\theta})\ket{\phi_j^{(0)}}$, which is also used to evaluate the cost function. With $H_{\rm eff}$, one can use the VQE approach to compute the low-energy spectra. However, considering the case that the dimension of $H_{\rm eff}$ is far less than $H$, we may be able to efficiently diagnolize $H_{\rm eff}$ to obtain the spectrum. In this work we simply diagnolize $H_{\rm eff}$ using a classical computer.

\begin{figure}[htbp]
\includegraphics[width = 0.5\textwidth]{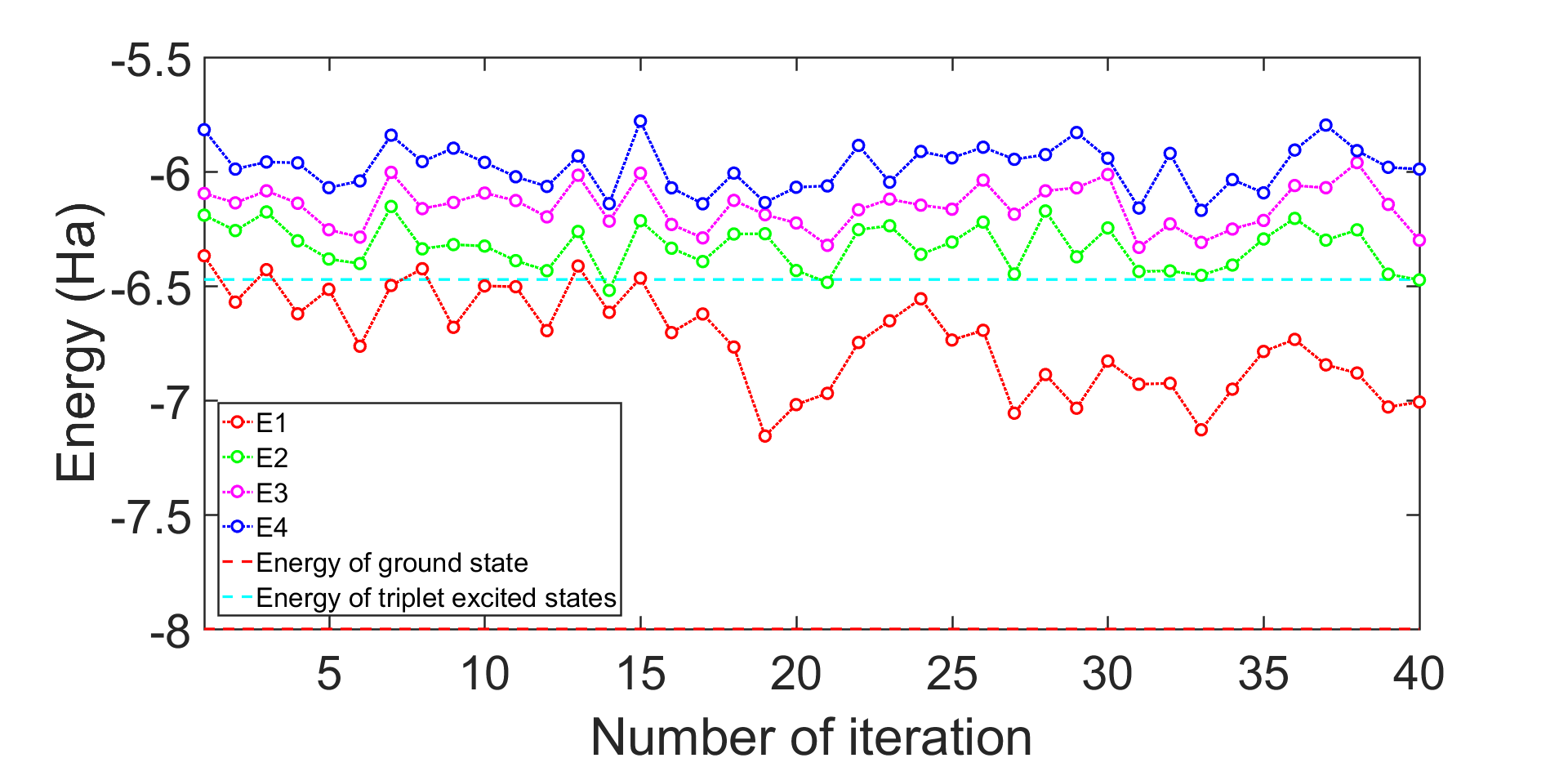}
\caption{\label{experimentally}Eigenvalues of $H_{\rm eff}$ experimentally measured using experimentally obtained parameters.}
\end{figure}

\begin{figure}[htbp]
\includegraphics[width = 0.5\textwidth]{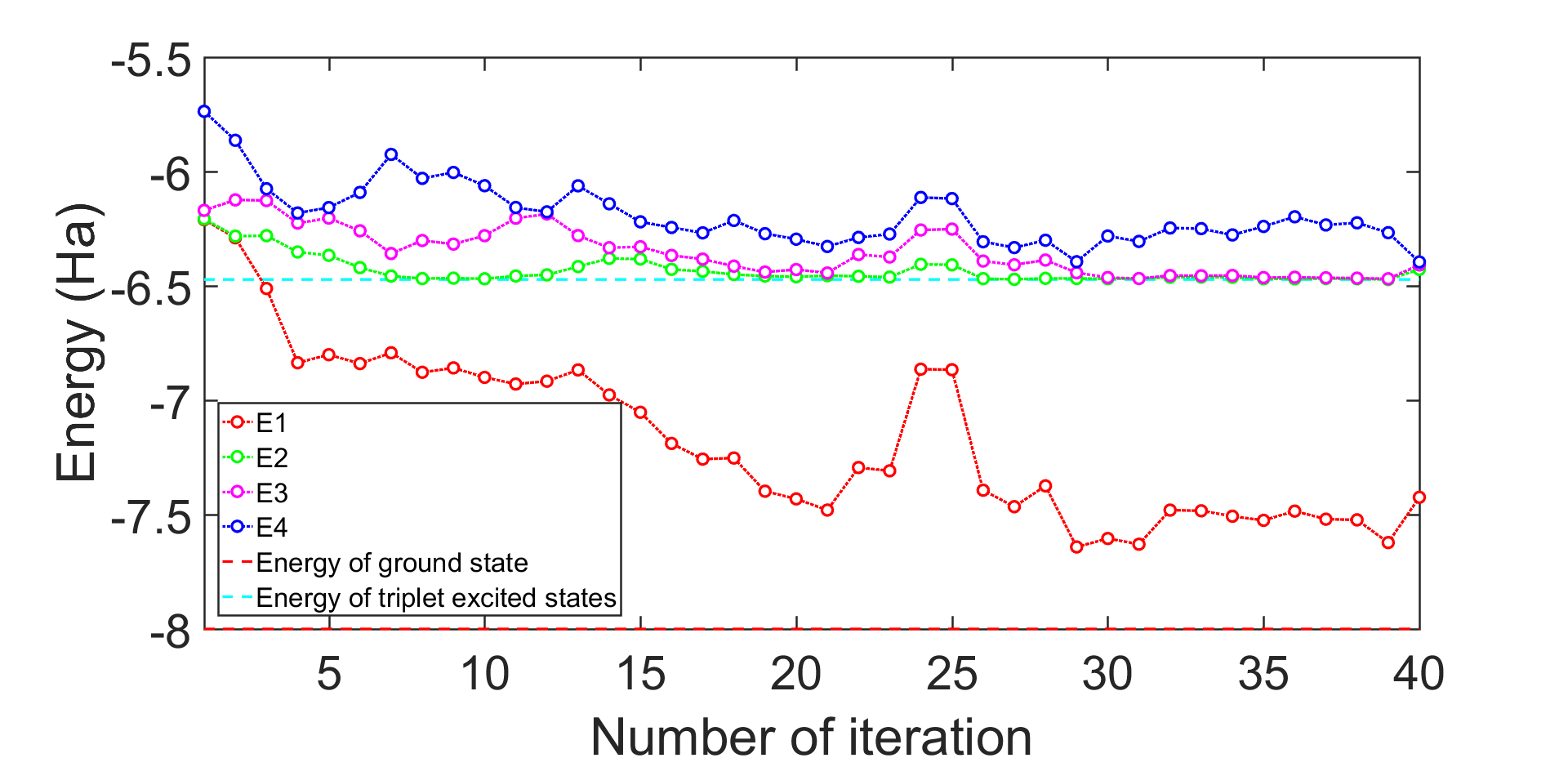}
\caption{\label{classically}Eigenvalues of $H_{\rm eff}$ classically computed using experimentally obtained parameters.}
\end{figure}

The results are shown in Fig.~\ref{experimentally}. As the iteration goes, the energies of the four eigenstates gradually approach the correct values with significant fluctuations. This is largely attributed to the high read out error of the quantum device, as can be supported by Fig.~\ref{classically}, where we classically compute the effective energy using the same set of parameters. We see smoother curves and the results are more close to the exact values. 

\begin{figure*}[htbp]
\includegraphics[width = 0.8\textwidth]{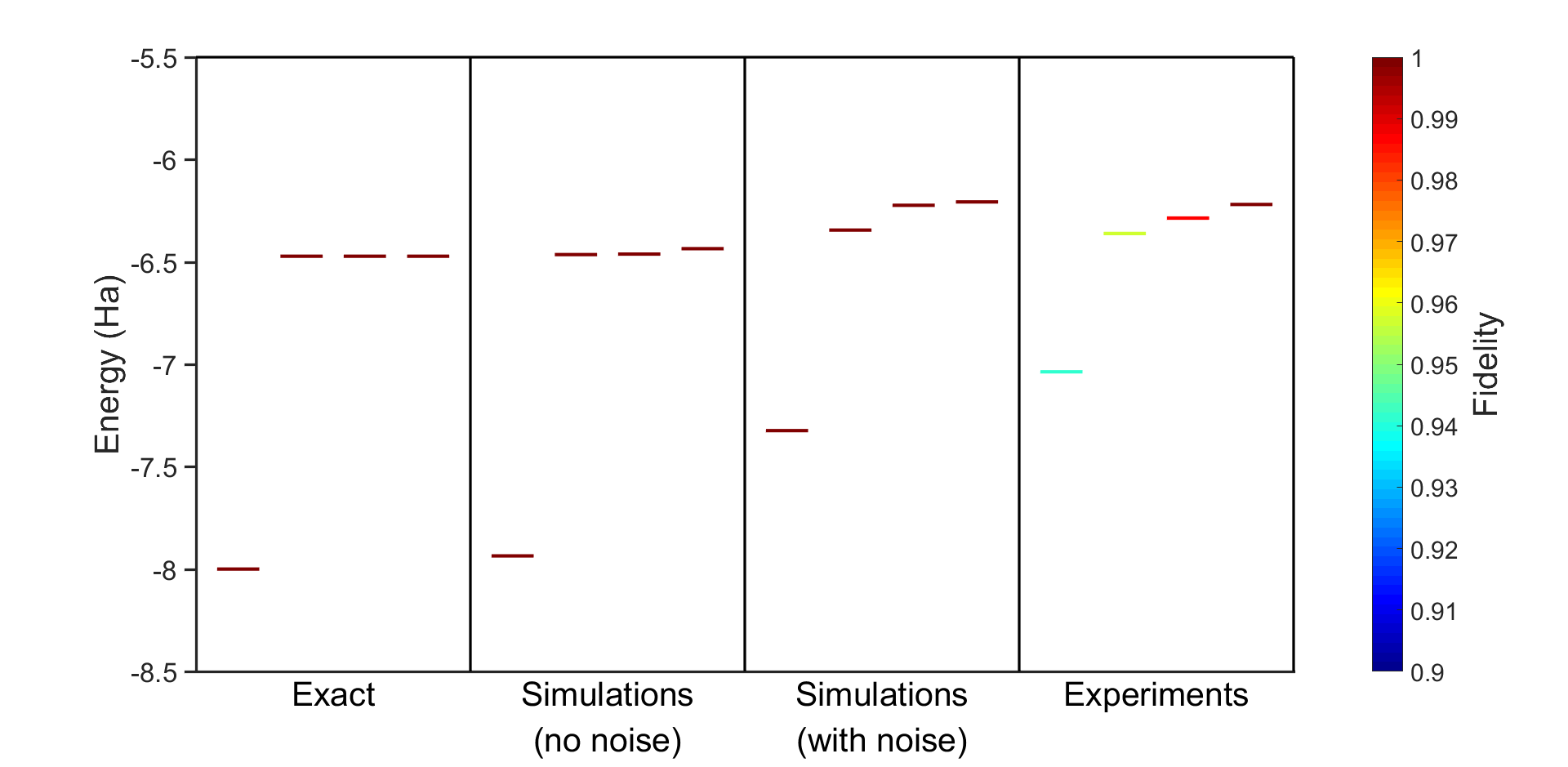}
\caption{\label{energy spectrum}The final energies in the low-energy spectrum and the fidelity of the obtained states with respect to the exact solutions. From left to right: exact energies, results from simulations with no noise, simulations with noise model from ibmq\_manila and experiments on ibmq\_manila. Horizontal solid line shows the energy value, with color indicating the fidelity with respect to the corresponding reference state.}
\end{figure*}

We summarize the results in Fig.~\ref{energy spectrum}, where we show the final energies obtained from both numerical simulations and experiments. The numerical simulations on the noise-free case were conducted using QuESTlink~\cite{jones2020questlink}, a packaged quantum emulator, and a noisy case was considered and simulated using ${\it qasm\_simualtor}$. The fidelity of the low-energy states with respect to the exact reference state, calculated by diagonalization of $H_{\rm eff}$ in each case, is reflected as the color of each short line. 

One thing needs to be noted is that to reduce the circuit depth, we use a simple ansatz which can well approximates the exact reference states but not a perfect choice. As can be seen from the second column in Fig.~\ref{energy spectrum}, there is a small gap between the exact solutions and the simulated results. Better results are expected with a more complex ansatz. 

Besides, the error mitigation technique relies on a good estimation of the noise. The CDR approach estimates the error information by measuring a set of observable using a circuit very close to the original one but with only Clifford gates. On IBMQ, circuits are running in batches, so we performed CDR circuits once in each batch. During the experiment, however, error fluctuates and deviates from the calibrated data and thus weakens the effect of error mitigation. The third column in Fig.~\ref{energy spectrum} shows the final energies obtained from numerical simulations on ${\it qasm\_simualtor}$ with a simplified noise model generated from the real-time calibration information of the ${\it ibmq\_manila}$ device (a function provided by IBMQ). Error mitigation was performed where the error information comes directly from the real-time calibration data. We see that with a better description of the error model, error mitigation performs better and the final results are more close to the exact values.

\section{\label{Conclusion}Discussion and Conclusion}

In this work, we for the first time proposed two quantum algorithms to realize the Schrieffer–Wolff transformation. The first algorithm constructs the SWT using a quantum circuit and evaluates the effective Hamiltonian with projection measurement. This method scales polynomially with the problem size and is suitable for fault-tolerant quantum computers. The second algorithm is a hybrid algorithm applicable for NISQ devices. This method is based on the variational algorithm, but instead of constructing a cost function to minimize the energy, our cost function is derived by using the block diagonalization property of the Hamiltonian. By optimizing the parameters, the cost function gradually approaches zero and the effective Hamiltonian can be obtained directly from the elements of the cost function. To verify this algorithm, we implemented it numerically and on an IBM quantum device using the example of a Heisenberg chain model with long-distance entanglement. The simulated results are very close to the exact values, and on a noisy device, we obtain the final states with over 95\% fidelity. To improve the results, some delicately designed error mitigation technique can be applied. 

Note that this method is not limited to find the ground state energy of a many-body system, but is able to find any energy interval of interest as long as they have no energy level crossings with others under perturbation.
See Appendix~\ref{Appendix general} for generalized description.

\begin{acknowledgments}
We thank Dayue Qin for discussions. We acknowledge
the use of simulation toolkit QuESTlink~\cite{jones2020questlink} and IBM Quantum services~\cite{IBMQ} for this work. We acknowledge the support of the National Natural Science Foundation of China (Grants No. 11875050 and No. 12088101) and NSAF (Grant No. U1930403).
\end{acknowledgments}

\bibliography{Quantum_algorithms_for_Schrieffer__Wolff_transformation}
\clearpage
\appendix
\section{\label{Appendix details}Details of section~\ref{first_algorithm}}

We explain the detailed steps sketched in Eq.(\ref{eq:CirRp}) in the main text as follows. Eq.(\ref{eq:CirRp}) involves steps 1-3, and the whole procedure to realize $R_{\mathscr{P}}\ket{\Psi}$ has in total 5 steps:
\begin{enumerate}
\item Initialize a $l$-qubit ancilla register in $\ket{+}^{\otimes l}$.
\item Apply controlled $e^{-iHt}$ from each ancilla qubit to the data qubits.
\item Apply inverse Fourier transform to the ancilla qubits. That completes a QPE cycle.
\item Apply the unitary gate $\sum_{k=0}^{k_{th}-1}\ketbra{k}{k} - \sum_{k=k_{th}}^{2^l-1}\ketbra{k}{k}$ to ancilla qubits.
\item Apply a reversed QPE to switch the state of the ancilla qubits back to $\ket{+}^{\otimes l}$.
\end{enumerate}

With similar steps, we can realize $R_{\mathscr{P}_0}R_{\mathscr{P}}\ket{\Psi}$, which will then be used as the controlled gate in Eq.(\ref{eq:CirU}).

The detailed steps of realizing SWT can be explained as follows. Eq.(\ref{eq:CirU}) includes the first three steps, and the whole procedure also has 5 steps:
\begin{enumerate}
\item Initialize a $l$-qubit ancilla register in $\ket{+}^{\otimes l}$.
\item Apply controlled $R_{\mathscr{P}_0}R_{\mathscr{P}}$ from each ancilla qubit to the data qubits.
\item Apply inverse Fourier transform to the ancilla qubits. That completes a QPE cycle.
\item Apply the phase gate $\sum_{k=0}^{2^l-1}e^{-i\frac{\pi k}{2^l}}\ketbra{k}{k}$ to ancilla qubits.
\item Apply a reversed QPE to switch the state of the ancilla qubits back to $\ket{+}^{\otimes l}$. Finally, the data qubits are in the state of $U\ket{\Psi}$.
\end{enumerate}

The circuit of the whole procedure is shown in Fig.~\ref{fig:structure}.
\begin{widetext}

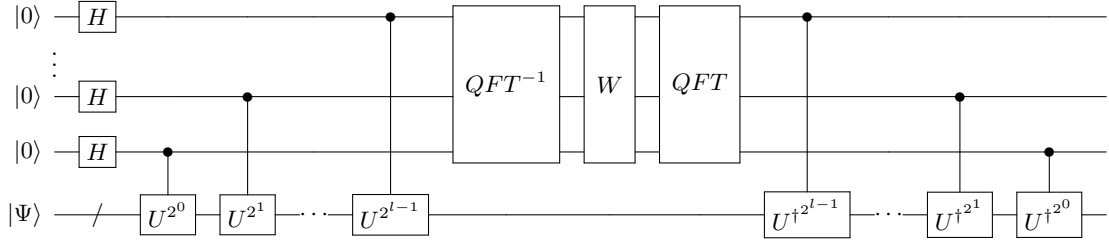
\begin{figure}[htbp]
\begin{tabular}{c}

\Qcircuit @C=1em @R=1em {
\lstick{\ket{0}}      & \gate{H}            & \qw                 & \qw    
& \qw                 &\ctrl{4}             & \multigate{3}{QFT^{-1}}  & \multigate{3}{W}
& \multigate{3}{QFT}&\ctrl{4}          & \qw                 & \qw 
& \qw                 & \qw\\
\vdots \\
\lstick{\ket{0}}      & \gate{H}            & \qw                 & \ctrl{2}       
& \qw                 & \qw                 & \ghost{QFT^{-1}}         & \ghost{W}   
& \ghost{QFT}  & \qw                 & \qw                 & \ctrl{2}
& \qw                 & \qw\\
\lstick{\ket{0}}      & \gate{H}            & \ctrl{1}            & \qw             
& \qw                 & \qw                 & \ghost{QFT^{-1}}         & \ghost{W}       
& \ghost{QFT}  & \qw                 & \qw                 & \qw
& \ctrl{1}            & \qw\\
\lstick{\ket{\Psi}}   & {/} \qw             & \gate{U^{2^0}}      & \gate{U^{2^1}} 
& \push{\cdots} \qw   & \gate{U^{2^{l-1}}}  & \qw                 & \qw
& \qw                 & \gate{U^{\dag^{2^{l-1}}}}&\push{\cdots} \qw & \gate{U^{\dag^{2^1}}}
& \gate{U^{\dag^{2^0}}} & \qw
}

\end{tabular}
\caption{\label{fig:structure}The general circuit for $R_{\mathscr{P}}$ and SWT. For the first case, $U=e^{-iHt}$. The $W$ gate makes a phase flip for those $k\geq k_{th}$ ancilla qubits states $\ket{k}$. For the second case, $U=R_{\mathscr{P}_0}R_{\mathscr{P}}$. The $W$ gate applies a conditional phase $e^{-i\frac{\pi k}{2^l}}$ to $\ket{k}$.}
\end{figure}
\end{widetext}

By the way, we show the quantum circuit used in our hybrid quantum-classical algorithm in Fig.~\ref{fig:ansatz}.

\begin{widetext}

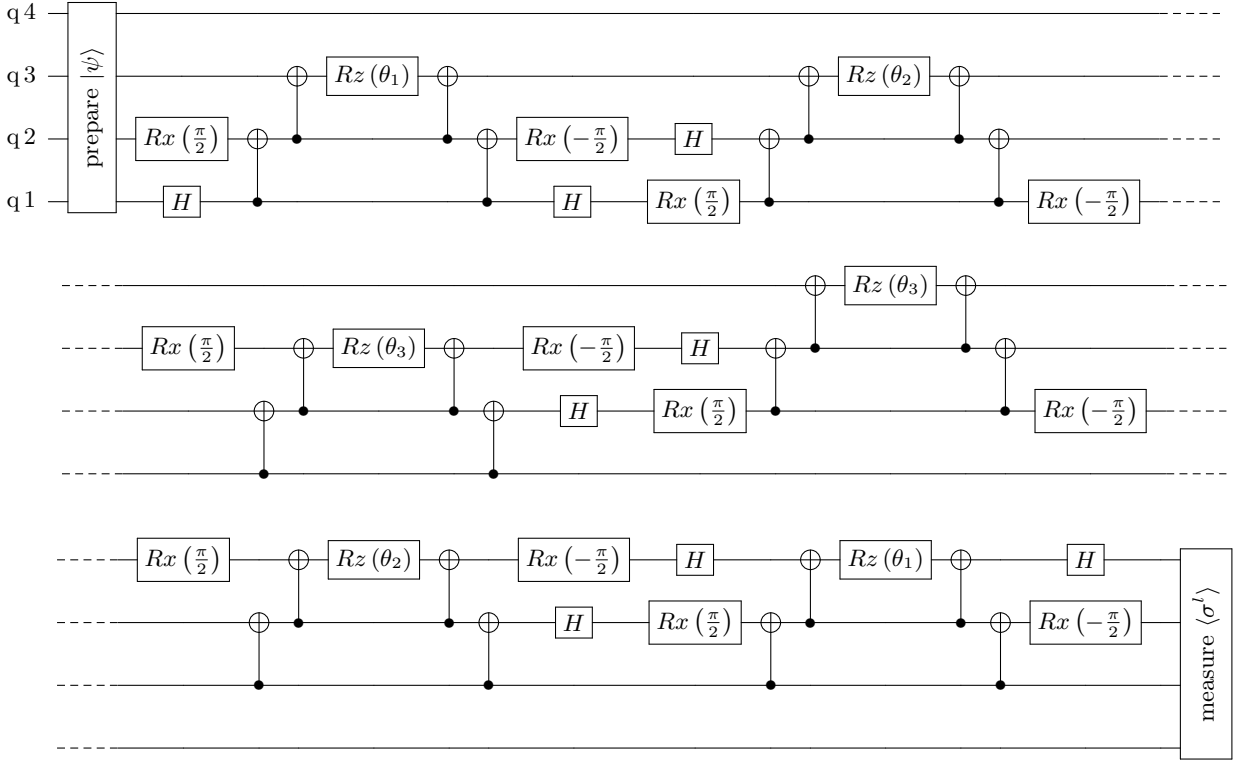
\begin{figure}[htbp]
\begin{tabular}{c}
\Qcircuit @C=0.8em @R=0.8em @!R {
\lstick{\rm{q}\,4}& \multigate{3}{\begin{turn}{90}prepare $\ket{\psi}$\end{turn}} & \qw & \qw & \qw & \qw & \qw & \qw & \qw & 
                   \qw & \qw & \qw & \qw & \qw & \qw & \qw   &\push{\begin{tikzpicture}
\draw[densely dashed](0,0)--(0.8,0);\end{tikzpicture}
}\qw\\
\lstick{\rm{q}\,3}& \ghost{\begin{turn}{90}prepare $\ket{\psi}$\end{turn}}& \qw & \qw & \targ & \gate{Rz\left(\theta_1\right)} & \targ & \qw & \qw & 
                   \qw & \qw & \targ & \gate{Rz\left(\theta_2\right)} & \targ & \qw & \qw &\push{\begin{tikzpicture}
\draw[densely dashed](0,0)--(0.8,0);\end{tikzpicture}}\qw\\
\lstick{\rm{q}\,2}& \ghost{\begin{turn}{90}prepare $\ket{\psi}$\end{turn}} & \gate{Rx\left(\frac{\pi}{2}\right)} & \targ & \ctrl{-1} & \qw & \ctrl{-1} & \targ & \gate{Rx\left(-\frac{\pi}{2}\right)} & 
                   \gate{H} & \targ & \ctrl{-1} & \qw & \ctrl{-1} & \targ & \qw &\push{\begin{tikzpicture}
\draw[densely dashed](0,0)--(0.8,0);\end{tikzpicture}}\qw\\
\lstick{\rm{q}\,1}& \ghost{\begin{turn}{90}prepare $\ket{\psi}$\end{turn}} & \gate{H} & \ctrl{-1} & \qw & \qw & \qw & \ctrl{-1} & \gate{H} &                           \gate{Rx\left(\frac{\pi}{2}\right)} & \ctrl{-1} & \qw & \qw & \qw & \ctrl{-1} & \gate{Rx\left(-\frac{\pi}{2}\right)} &\push{\begin{tikzpicture}
\draw[densely dashed](0,0)--(0.8,0);\end{tikzpicture}}\qw\\
}

\\[10em]

\Qcircuit @C=0.8em @R=0.8em @!R {
&\push{\begin{tikzpicture}
\draw[densely dashed](0,0)--(0.8,0);\end{tikzpicture}} & \qw & \qw & \qw & \qw & \qw & \qw & \qw & 
                    \qw & \qw & \targ & \gate{Rz\left(\theta_3\right)} & \targ & \qw & \qw &\push{\begin{tikzpicture}
\draw[densely dashed](0,0)--(0.8,0);\end{tikzpicture}}\qw\\
&\push{\begin{tikzpicture}
\draw[densely dashed](0,0)--(0.8,0);\end{tikzpicture}} & \gate{Rx\left(\frac{\pi}{2}\right)} & \qw & \targ & \gate{Rz\left(\theta_3\right)} & \targ & \qw & \gate{Rx\left(-\frac{\pi}{2}\right)} & 
                   \gate{H} & \targ & \ctrl{-1} & \qw & \ctrl{-1} & \targ & \qw &\push{\begin{tikzpicture}
\draw[densely dashed](0,0)--(0.8,0);\end{tikzpicture}}\qw\\
&\push{\begin{tikzpicture}
\draw[densely dashed](0,0)--(0.8,0);\end{tikzpicture}} & \qw & \targ & \ctrl{-1} & \qw & \ctrl{-1} & \targ & \gate{H} & 
                   \gate{Rx\left(\frac{\pi}{2}\right)} & \ctrl{-1} & \qw & \qw & \qw & \ctrl{-1} & \gate{Rx\left(-\frac{\pi}{2}\right)} &\push{\begin{tikzpicture}
\draw[densely dashed](0,0)--(0.8,0);\end{tikzpicture}}\qw\\
&\push{\begin{tikzpicture}
\draw[densely dashed](0,0)--(0.8,0);\end{tikzpicture}} & \qw & \ctrl{-1} & \qw & \qw & \qw & \ctrl{-1} & \qw & 
                   \qw & \qw & \qw & \qw & \qw & \qw & \qw &\push{\begin{tikzpicture}
\draw[densely dashed](0,0)--(0.8,0);\end{tikzpicture}}\qw\\
}                   

\\[10em]

\Qcircuit @C=0.8em @R=0.8em @!R {                   
&\push{\begin{tikzpicture}
\draw[densely dashed](0,0)--(0.8,0);\end{tikzpicture}} & \gate{Rx\left(\frac{\pi}{2}\right)} & \qw & \targ & \gate{Rz\left(\theta_2\right)} & \targ & \qw & \gate{Rx\left(-\frac{\pi}{2}\right)} & 
                   \gate{H} & \qw & \targ & \gate{Rz\left(\theta_1\right)} & \targ & \qw & \gate{H} & \qw & \multigate{3}{\begin{turn}{90}measure $\mean{\sigma^l}$\end{turn}}\\                 
&\push{\begin{tikzpicture}
\draw[densely dashed](0,0)--(0.8,0);\end{tikzpicture}} & \qw & \targ & \ctrl{-1} & \qw & \ctrl{-1} & \targ & \gate{H} & 
                   \gate{Rx\left(\frac{\pi}{2}\right)} & \targ & \ctrl{-1} & \qw & \ctrl{-1} & \targ & \gate{Rx\left(-\frac{\pi}{2}\right)} & \qw & \ghost{\begin{turn}{90}measure $\mean{\sigma^l}$\end{turn}}\\                   
&\push{\begin{tikzpicture}
\draw[densely dashed](0,0)--(0.8,0);\end{tikzpicture}} & \qw & \ctrl{-1} & \qw & \qw & \qw & \ctrl{-1} & \qw & 
                   \qw & \ctrl{-1} & \qw & \qw & \qw & \ctrl{-1} & \qw & \qw & \ghost{\begin{turn}{90}measure $\mean{\sigma^l}$\end{turn}}\\
&\push{\begin{tikzpicture}
\draw[densely dashed](0,0)--(0.8,0);\end{tikzpicture}} & \qw & \qw & \qw & \qw & \qw & \qw & \qw & 
                  \qw & \qw & \qw & \qw & \qw & \qw & \qw & \qw & \ghost{\begin{turn}{90}measure $\mean{\sigma^l}$\end{turn}}\\
}
\end{tabular}
\caption{\label{fig:ansatz}The quantum circuit used in our experiments. In preparation, the state $\ket{\psi}$ is $\ket{\phi_i^{(0)}}$ or $\frac{I\pm G}{2}\ket{\phi_i^{(0)}}$ (up to a normalization factor). In measurement, we measure the expectation of $\sigma^l$. The ansatz circuit is between these two blocks.}
\end{figure}

\end{widetext}

\section{\label{Appendix derivation}Derivation of coefficients}
The Taylor series of $e^{-iHt}$ is
\begin{eqnarray}
I-iHt-\frac{1}{2}H^2t^2+O(t^3) .
\end{eqnarray}
Substitute it into Eq.~(\ref{eq:evolution}) yields
\begin{widetext}
\begin{align}
\mathcal{L}_t=&-\frac{1}{M}\sum_{i, j}^M\left|\bra{\phi_i^{(0)}}U (I-iHt-\frac{1}{2}H^2t^2+O(t^3)) U^\dag\ket{\phi_j^{(0)}}\right|^2 \notag\\
=&-\frac{1}{M}\sum_{i=j}^M\left| 1-it\bra{\phi_i^{(0)}}UHU^\dag\ket{\phi_i^{(0)}}-\frac{1}{2}t^2\bra{\phi_i^{(0)}}UH^2U^\dag\ket{\phi_i^{(0)}}+O(t^3) \right|^2 \notag\\
&-\frac{1}{M}\sum_{i \neq j}^M\left|  -it\bra{\phi_i^{(0)}}UHU^\dag\ket{\phi_j^{(0)}}-\frac{1}{2}t^2\bra{\phi_i^{(0)}}UH^2U^\dag\ket{\phi_j^{(0)}}+O(t^3) \right|^2 \notag\\
=&-\frac{1}{M}\sum_{i=j}^M\left[(1-\frac{1}{2}t^2\bra{\phi_i^{(0)}}UH^2U^\dag\ket{\phi_i^{(0)}})^2+t^2\bra{\phi_i^{(0)}}UHU^\dag\ket{\phi_i^{(0)}}^2\right] \notag\\
&-\frac{1}{M}\sum_{i \neq j}^M\left| -it\bra{\phi_i^{(0)}}UHU^\dag\ket{\phi_j^{(0)}} \right|^2+O(t^3) \notag\\
=&-\frac{1}{M}\sum_{i=j}^M(1-t^2\bra{\phi_i^{(0)}}UH^2U^\dag\ket{\phi_i^{(0)}}+t^2\bra{\phi_i^{(0)}}UHU^\dag\ket{\phi_i^{(0)}}^2) \notag\\
&-\frac{1}{M}\sum_{i \neq j}^M t^2\left| \bra{\phi_i^{(0)}}UHU^\dag\ket{\phi_j^{(0)}} \right|^2+O(t^3) \notag\\
=&-1+\frac{t^2}{M}(\sum_{i=1}^M\bra{\phi_i^{(0)}}UH^2U^\dag\ket{\phi_i^{(0)}}-\sum_{i, j}^M\left| \bra{\phi_i^{(0)}}UHU^\dag\ket{\phi_j^{(0)}} \right|^2)+O(t^3) .
\end{align}
\end{widetext}

\section{\label{Appendix equivalence}Equivalence}
Recall that $P$ is the projector in low-energy subspace $\mathscr{P}$ of $H$, $Q$ is its orthogonal complement which represents the high-energy subspace projector. They can be written as
\begin{eqnarray}
P=\sum_{i=1}^M U^\dag\ket{\phi_i^{(0)}}\bra{\phi_i^{(0)}}U=\sum_{i=1}^M \ket{\phi_i}\bra{\phi_i} , 
\label{eq:P}
\end{eqnarray}
\begin{eqnarray}
Q=\sum_{i=M+1}^N U^\dag\ket{\phi_i^{(0)}}\bra{\phi_i^{(0)}}U=\sum_{i=M+1}^N \ket{\phi_i}\bra{\phi_i} .   
\end{eqnarray}
With Eq.~(\ref{eq:P}), $C$ can be converted to
\begin{eqnarray}
C&=&\sum_{i=1}^M \bra{\phi_i}H^2\ket{\phi_i}-\sum_{i, j}^M\left| \bra{\phi_i}H\ket{\phi_j} \right|^2 \notag\\
&=&\sum_{i=1}^M \bra{\phi_i}H^2\ket{\phi_i}-\sum_{i, j}^M\bra{\phi_i}H\ket{\phi_j}\bra{\phi_j}H\ket{\phi_i} \notag\\
&=&Tr(PH^2P)-Tr(PHPHP) ,
\label{eq:C}
\end{eqnarray}
where the trace operation acts on the Hilbert space of whole system.

Define
\begin{eqnarray}
H_P=PHP ,
\label{eq:H_P}
\end{eqnarray}
\begin{eqnarray}
H_Q=QHQ ,
\end{eqnarray}
\begin{eqnarray}
H_{PQ}=PHQ ,
\end{eqnarray}
\begin{eqnarray}
H_{QP}=QHP .
\end{eqnarray}
The Hamiltonian can be devided into
\begin{eqnarray}
H=H_P+H_Q+H_{PQ}+H_{QP},   
\end{eqnarray}
and its square is
\begin{align}
H^2&=H_P^2+H_PH_{PQ}+H_Q^2+H_QH_{QP} \notag \\
&+H_{PQ}H_Q+H_{PQ}H_{QP}+H_{QP}H_P+H_{QP}H_{PQ} . 
\end{align}
So we have
\begin{eqnarray}
PH^2P=H_P^2+H_{PQ}H_{QP} .
\label{eq:PH^2P}   
\end{eqnarray}

Substitute Eq.~(\ref{eq:H_P}) and Eq.~(\ref{eq:PH^2P}) into Eq.~(\ref{eq:C}), we get
\begin{eqnarray}
C&=&Tr(PH^2P)-Tr(PHPPHP) \notag\\
&=&Tr(H_P^2)+Tr(H_{PQ}H_{QP})-Tr(H_P^2) \notag\\
&=&Tr(H_{PQ}H_{QP}) .
\end{eqnarray}
Therefore, $C=0$ if and only if $H_{PQ}$ is zero matrix which indicates $H$ is completely block-diagonal with respect to $P$ and $Q$.

\section{\label{Appendix general}General case of SWT}
In a general situation, the unperturbed Hamiltonian is
\begin{align}
H_0 &= \sum_{i=1}^K E_i^{(0)} \ket{\phi_i^{(0)}}\bra{\phi_i^{(0)}} + \sum_{i=K+1}^{K+M} E_i^{(0)} \ket{\phi_i^{(0)}}\bra{\phi_i^{(0)}}\\
&+ \sum_{i=K+M+1}^N E_i^{(0)} \ket{\phi_i^{(0)}}\bra{\phi_i^{(0)}},
\end{align}
with eigenvalues $E_1^{(0)} \leq \cdots \leq E_K^{(0)} < E_{K+1}^{(0)} \leq \cdots \leq E_{K+M}^{(0)} < E_{K+M+1}^{(0)} \leq \cdots \leq E_N^{(0)}$. The projector onto the eigenspace of eigenvalues $E_i^{(0)} \in [E_{K+1}^{(0)}, E_{K+M}^{(0)}]$ reads
\begin{equation}
P_0 = \sum_{i=K+1}^{K+M} \ket{\phi_i^{(0)}}\bra{\phi_i^{(0)}}.
\end{equation}
Here the energy gap is 
\begin{eqnarray}
\Delta = {\rm min} \{ E_{K+1}^{(0)} - E_K^{(0)}, E_{K+M+1}^{(0)} - E_{K+M}^{(0)} \}.
\end{eqnarray}
Through the same assumption Eq.(\ref{eq:shift}), the total Hamiltonian can be written as
\begin{align}
H &= \sum_{i=1}^K E_i \ket{\phi_i}\bra{\phi_i} + \sum_{i=K+1}^{K+M} E_i \ket{\phi_i}\bra{\phi_i}\\
&+ \sum_{i=K+M+1}^N E_i \ket{\phi_i}\bra{\phi_i},   
\end{align}
with eigenvalues $E_1 \leq \cdots \leq E_K < E_{K+1} \leq \cdots \leq E_{K+M} < E_{K+M+1} \leq \cdots \leq E_N$. The eigenspace $\mathscr{P}$ is spanned by $H$'s eigenstates with eigenvalues $E_i \in ( E_{K+1}^{(0)} - \Delta/2, E_{K+M}^{(0)} + \Delta/2 )$. Then the effective energy spectra, from $E_{K+1}$ to $E_{K+M}$, can be obtained following the same method.

\end{document}